\documentstyle[12pt,nature,epsf]{article}

\begin{document}
\newcommand{\be}{\begin{eqnarray}}
\newcommand{\ee}{\end{eqnarray}}
\newcommand{\etal}{{\it{et al.}}}
\newcommand{\smass}{M_{\odot}}
\newcommand{\br}{{\bf r}}
\newcommand{\bV}{{{\bf v}}}


%
%
\title{
Kuiper-belt Binary Formation through Exchange Reactions
}

%
%

\author{Yoko Funato\\
General System Studies,
University of Tokyo,
    Tokyo 153, Japan\\
\medskip
\small{\tt funato@chianti.c.u-tokyo.ac.jp}\\
Junichiro Makino\\
\bigskip
Department of Astronomy, University of Tokyo, Tokyo 113, Japan\\
%
Piet Hut\\
\bigskip
Institute for Advanced Study,
Princeton, NJ 08540, USA\\
%
Eiichiro Kokubo \& Daisuke Kinoshita\\
National Astronomical Observatory, Tokyo 180, Japan}

%
%

\def\jun#1{{\bf[#1---Jun]}}
\def\piet#1{{\bf[#1---Piet]}}
\def\yoko#1{{\bf[#1---Yoko]}}

\maketitle

\begin{abstract}

{\bf
Recent observations\cite{Burnes2002,Veillet2002,Margot2002a} have
revealed an unexpectedly high binary fraction among the
Trans-Neptunian Objects (TNOs) that populate the Kuiper Belt.  The TNO
binaries are strikingly different from asteroid binaries in four
respects\cite{Veillet2002}: their frequency is an order of magnitude
larger, the mass ratio of their components is closer to unity, and
their orbits are wider and highly eccentric.
Two explanations have been proposed for their formation, one assuming
large numbers of massive bodies\cite{Weidenschilling2002}, and
one assuming large numbers of light bodies\cite{Goldreich2002}.  We
argue that both assumptions are unwarranted, and we show how TNO
binaries can be produced from a modest number of intermediate-mass
bodies of the type predicted by the gravitational instability theory
for the formation of planetesimals\cite{GoldreichWard1973}.
We start with a TNO binary population similar to the asteroid binary
population, but subsequently modified by three-body exchange
reactions, a process that is far more efficient in the Kuiper belt,
because of the much smaller tidal perturbations by the Sun.  Our
mechanism can naturally account for all four characteristics that
distinguish TNO binaries from main-belt asteroid binaries.
}
\end{abstract}

\def\simlt{\hbox{ \rlap{\raise 0.425ex\hbox{$<$}}\lower 0.65ex
  \hbox{$\sim$} }}
\def\simgt{\hbox{ \rlap{\raise 0.425ex\hbox{$>$}}\lower 0.65ex
  \hbox{$\sim$} }} 

The TNO binary 1998WW31 has the following dynamical
properties\cite{Veillet2002}: mass ratio $m_2/m_1 \sim 0.7$,
eccentricity $e \sim 0.8$, semimajor axis $a \sim 2\times 10^4$ km,
and inferred radii $r_1 \sim 1.1r_2 \sim 10^2$ km, where $m_1(r_1)$
and $m_2(r_2)$ are the masses (radii) of the primary and the secondary,
hence $a/r_1 > 10^2$.  All this stands in stark contrast to the
properties of typical main belt asteroid binaries\cite{Margot2002b},
where $m_2/m_1 \ll 1$, $e \sim 0$, and $a/r_1 \simlt 10$.

Asteroid binaries are thought to have been formed by
collisions\cite{Merline2003}, in a scenario similar to the leading
candidate for the formation of the Earth-Moon
system\cite{HartmannDavis1975,Kokubo2000}: two asteroids collide,
leaving a small fraction of their combined matter with a large
fraction of their relative angular momentum in a disk.  Some of the
disk matter then quickly coagulates into a small companion.  The
observed characteristics, $m_2/m_1 \ll 1$, $e \sim 0$, and
$a/r_1 \simlt 10$, are all natural consequences of this
scenario\cite{Durda2001}.

Clearly, we need a different mechanism for the formation of 1998WW31.
We can look for an analogy by considering dynamical binary formation
among stars in dense stellar systems, where there are three channels:
1) tidal capture, where tidal dissipation during a close encounter
between two single stars leaves the system bound\cite{FPR1975};
2) three-body binary formation, in a simultaneous close encounter 
between three single stars where one of the three stars carries the
excess energy away, leaving the other two stars bound\cite{Heggie1975};
and 3) exchange reactions, where a single star encounters a binary,
and replaces one of its original members\cite{Heggie1975}.

Channel 1 is analogous to the standard scenario for asteroid binary
formation.  It will indeed occur: each TNO has grown through
accretion, and much of this accretion has happened through collisions
with an object comparable in mass to the growing TNO
itself\cite{Makino1998}; in some of these collisions there will be too
much angular momentum to form a single stable body, and in these cases
it is unavoidable to form tight circular
binaries of strongly unequal mass.  Such binaries, when formed as
intermediary stages during the growth of a TNO, are short-lived.  The cross
section for disruption of a binary is related to the cross section for
accretion by the ratio of the sizes of the binary orbit and the TNO
primary.  Therefore, the binary will be broken up by a perturbation of
a third body with mass comparable to the primary well before such a
body will hit the primary itself, in the vast majority of cases.

Channel 2 would require a near-simultaneous encounter of three massive
objects with low enough velocities to allow an appreciable chance to leave
two of the objects bound.  For this to work, the random velocities of
the most massive objects should be significantly lower than their
Hill velocities.  Under such conditions, this channel could play a
role, as pointed out by Goldreich {\it et al.}\cite{Goldreich2002}, 
who assumed that there are $\sim10^5$ 100 km--sized object embedded
in a sea of small ($<1 {\rm km}$) objects.  This assumption, however,
is at odds with Goldreich and Ward's theory for the formation of
planetesimals\cite{GoldreichWard1973} through gravitational instability,
and it is hard to see how objects in the Kuiper Belt could form from
non-gravitational coagulation, because the time scales are far too
long\cite{Wetherill1990}.  In contrast, the gravitational instability
theory predicts the size of the initial bodies to be $10-100 {\rm km}$.
Starting with these larger bodies would make channel 2 ineffective,
because the velocity dispersion would be higher than the Hill
velocity\cite{KokuboIda1997,KenyonLuu1998}.

Recently, Weidenschilling\cite{Weidenschilling2002} proposed a
variation on the idea of using interactions between three unbound
bodies in order to create a binary.  He studied how a third massive
body could capture the merger remnant from a collision of two massive
bodies if the third body were near enough during the time of the
collision.  This mechanism seems unlikely to work, however, since it
requires a number density of massive objects about two orders of
magnitude higher than the value consistent with present
observations\cite{Goldreich2002}.

Goldreich {\it et al.}\cite{Goldreich2002} have proposed another
mechanism, based on the dynamical friction from a sea of smaller
bodies that can turn a hyperbolic encounter between two massive
bodies into a bound orbit under favorable conditions.  Effectively,
this mechanism makes use of a superposition of three-body encounters,
since each light body interacts independently with the two heavier
ones, and in that sense it is another variant on channel 2.
As we mentioned above, the gravitational instability theory for the
formation of planetesimals\cite{GoldreichWard1973} would exclude the
existence of such a sea of small objects, and since the alternative
theory of nongravitational agglomeration does not seem to work, we
will explore the consequences of dropping channel 2, which leaves us
with channel 3.

Channel 3 can only operate when there are already binaries available
for encounters.  The only binaries that are expected to be formed
frequently are the ones produced in channel 1, so we should consider
channel 1 and 3 in tandem.  What remains to be done is to check
whether the resulting properties of the binaries produced are the ones
we are looking for in the Kuiper belt, and to check whether the
formation efficiency is high enough to explain the abundant presence
of surviving TNO binaries.

Starting with the first task, consider a relatively massive TNO
primary in a binary orbit with a much less massive secondary, embedded
in a sea of smaller particles, most of which are far less massive than
either of the binary components.  The smallest particles, when they
come close, will simply accrete on either the secondary or primary or
just pass through the system.  The cumulative effect may drive the
binary components to collide, which is fine, since then sooner or
later a large impact is likely to create again a new temporary binary
of a similar type.  However, if the binary encounters a particle with
a mass $m$ that is comparable to the mass of the primary
component ($m_1 \sim m \gg m_2$), the most likely result is an exchange
reaction, in which the incoming object replaces the original
secondary\cite{Spitzer1987}. Figure 1 shows an example of such a
reaction.

\begin{figure}
\begin{center}
\leavevmode
\epsfxsize 10cm
\epsffile{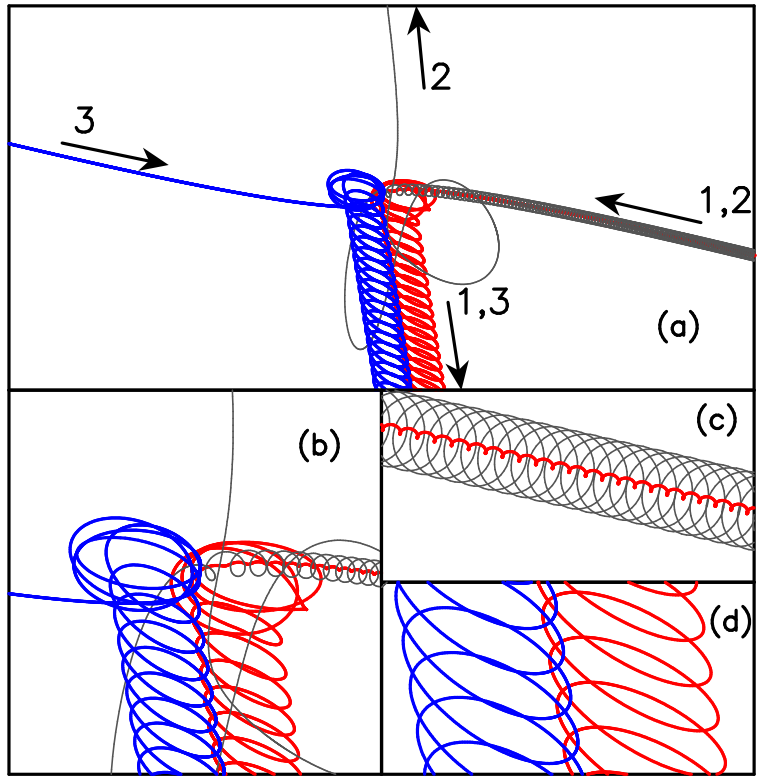}
\end{center}
\label{fig:1a}

\caption{
{\narrower
 An example of a binary--single-body exchange interaction, in the
`(massive, light) meets massive' category discussed in this paper.
Bodies 1 and 2 have masses $m_1=1$ and $m_2=0.1$, respectively,
forming a binary with an initially circular orbit.  Body 3, with mass
$m_3=1$, encounters the binary on an initially parabolic orbit.  In
panel (a), the whole scattering process is shown.  Panel (b) shows the
complex central interaction in more detail, while panels (c) and (d)
show the orbits of the initial and final binary, respectively.
Note that the final binary orbit is highly eccentric and much wider
than the initial circular binary orbit.
}
}
\end{figure}

Using the impulse approximation for the replacement, together with
conservation of energy and specific angular momentum, we can make a crude
estimate for the semimajor axis $a$ and eccentricity $e$ of the newly
formed binary in terms of the corresponding values $a_0$ and $e_0 = 0$
for the original binary.  Viewing the replacement as an isolated
two-body encounter, we see immediately that the asymptotic escape
speed of the initial secondary must be comparable to that of the
incoming velocity at infinity of the third body.  Since the latter is
much smaller than the relative velocity of the binary components (as
a necessary condition for runaway growth), the binding energy of the
binary will
not change much during the exchange, hence $m_1m_2/a_0 \approx m_1m/a$
where $a$ is the new semimajor axis after the exchange.  This implies
$a/a_0 \approx m/m_2 \gg 1$: the size of the orbit of the binary is
increased in proportion to the increase in mass of the object orbiting
the primary.  Under the impulse approximation, the interaction happens
in a space small compared to the distance $a_0$ to the primary.
Neglecting factors of $\simlt 2$ related to reduced mass, this implies
that the angular
momentum per unit mass of the newly bound particle with respect to the
primary is the same as that of the previous companion.  Conservation
of specific angular momentum of the system then gives $m_2a_0(1-e_0)
\approx ma(1-e)$ which gives $1-e \approx m_2/m \ll 1$: the
eccentricity of the new binary is almost unity, and the orbit is very
elongated.

While the impulse approximation may not be a very good assumption, at
least qualitatively it predicts that exchange reactions provide us
with just what we wanted: binaries with small mass ratios in large
highly eccentric orbits.  In order to perform a more quantitative
check, we have run a series of scattering experiments to obtain the
relevant cross sections.  Here, we assumed that the initial binary has
a mass ratio of $20:1$ and semimajor axis $a_0=20r_1$, where $r_1$ is
the radius of the primary.  These values are  typical for main-belt
binary asteroids, with $m_2/m_1 < 0.1$, and separations $5-40$ times
the radius of the primary.  We choose parabolic relative orbits for the
single body approaching the binary, with periastron distances uniformly
distributed between 0 and $20a_0$.  In case of resonance reactions,
where the incoming body is captured, and the whole system undergoes a
complicated three-body dance, we only followed the system as long as
all three bodies stayed within a maximum distance of $1000a_0$ from
the other two.  If this condition was violated for any of the three
bodies, we considered that body to escape, due to the perturbing tidal
field of the Sun, which in the Kuiper belt corresponds to a Hill
radius of order of $10^3a_0$.

Table 1 shows the cross sections for those processes in which the
initial binary membership is altered.  Channels (a), (c) and (e)
result in binaries with two massive components, while channels
(b) and (d) produce binaries with large mass ratios.  There are
four remaining channels that do not produce any binary: a triple
merger 1+2+3, and three channels in which two of the three bodies
merge while the third body escapes.  The cross sections of these four
cases are summed together under (f).  Note that about 80\% of
these interactions result in binaries with two massive components.
These cross sections were calculated using a scattering code that
incorporated an effective tidal cut-off, and the results were checked
independently through a comparison with the starlab three-body
scattering package\cite{McMH1996}.

\begin{table}
\label{tbl:CrossSection}
\begin{center}
\begin{tabular}{cccccccc}
\hline\hline
channel: & (a) & (b) & (c)  & (d) & (e) & (f) \\
\hline\hline
process: & (1,3),2  & (2,3),1 & (1+2,3) &
 (1+3,2)  & (2+3,1) & no binary \\
\hline
$\sigma v^2$: & 12.1 & 1.3 & 0.9 & 1.3 & 0.9 & 1.2\\
\hline
\end{tabular}
\end{center}
{
Table 1:
Cross sections $\sigma$ for various configuration-changing channels in
binary--single-body scattering.  The gravitational focusing factor
$v^2$ is scaled out in order to obtain finite values in the parabolic
limit, where $v$ is the initial relative velocity between binary and
single body at infinity.  We use units in which $G=m_1=m_3=a=1$,
where $G$ is the gravitational constant, $m_1$ and $m_3$ are the
masses of the heaviest body in the binary and the single body,
respectively, and $a$ is the initial semi-major axis of the binary.
The mass of the lighter body in the binary is $m_2=0.05$.
The radii are $r_1=r_3=0.05$ and $r_2=r_1(m_2/m_1)^{1/3}\approx 0.01842$.
The scattering processes are coded as follows: $(x,y)$ indicates a
binary in the final state with components $x$ and $y$, while a $p+q$
indicates the product of a merger between bodies $p$ and $q$.  A
single body $z$ in the final state is indicated by $(,),z$.
The physical meaning of the six channels is as follows:
(a) an exchange reaction resulting in a massive--massive binary;
(b) an exchange reaction resulting in a massive--light binary;
(c) a merger resulting in a massive--massive binary;
(d) a merger resulting in a twice-as-massive--light binary;
(e) a merger resulting in a massive--massive binary;
(f) no binary is left, after three-body merging or two-body merging
followed by escape.
}

\end{table}

Figure 2 shows the normalized differential cross sections for the
semimajor axis $a$ and eccentricity $e$, for forming a final binary
with two massive components. The distribution for the semi-major axis
is strongly peaked at $a=20$, in good agreement with the simple argument
presented above.  Similarly, the eccentricity peaks at 0.95, as expected.
For the physical values of the semi-major axis, we have assumed a radius
radius for the primary of $r_1=75 {\rm km}$, the estimated size of
the primary of 1998WW31, based on the mass deduced from the binary
motion and an assumed mean density of $1 {\rm g/cm^3}$.
Figure 3 shows the distribution of the final binary in the
$a,e$ plane.  The orbital elements of 1998WW31 are consistent with
the binary having formed through the processes modeled here.

\begin{figure}
\begin{center}
\leavevmode
\epsfxsize 10cm
\epsffile{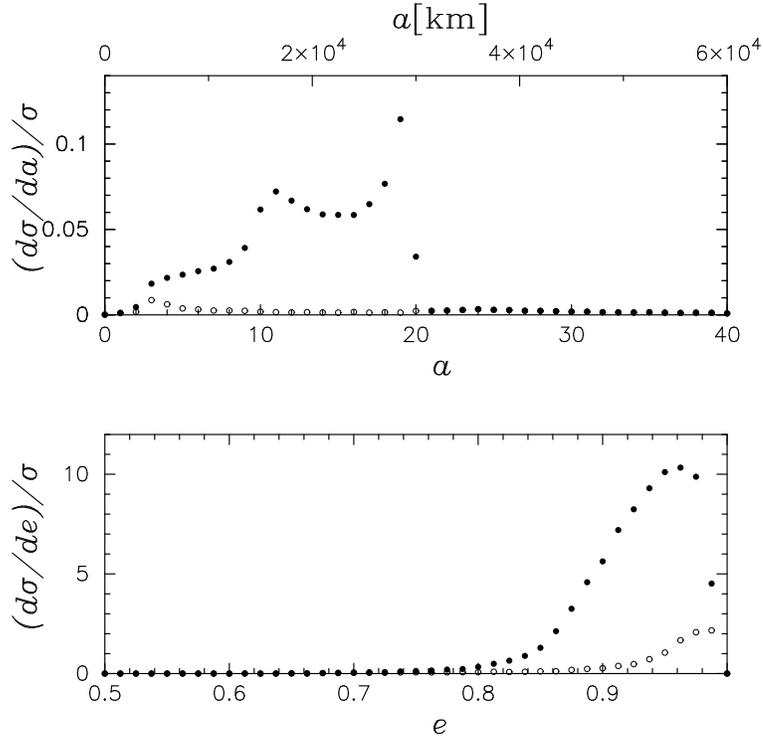}
\end{center}
\label{fig:2}

\caption{ Normalized differential cross sections for the formation of
a `massive-massive' binary, under the conditions specified in the text
(channels a, c and e in table 1), with respect to the semi-major axis
$a$ (top panel), and eccentricity $e$ (bottom panel) of the final
binary.  The initially circular binary has $a=1$ in the dimensionless
units used for $d\sigma/da$, while the physical units are given for
reference at the top of the figure.  The filled points are the total
values for the differential cross sections, while the open circles are
the contributions from the merger channels (c and e in table 1).  Note
the double-peaked structure in the top panel: the sharp peak toward
$a\sim 20$ arises from non-resonant exchanges, where the final binary
has an energy comparable to that of the initial binary; the broad peak
around $a\sim 10$ arises from resonant exchanges, where the memory of
the initial binary is wiped out, leading on average to more strongly
hyperbolic escape in which a harder binary is formed.  }
\end{figure}

\begin{figure}
\begin{center}
\leavevmode
\epsfxsize 10cm
\epsffile{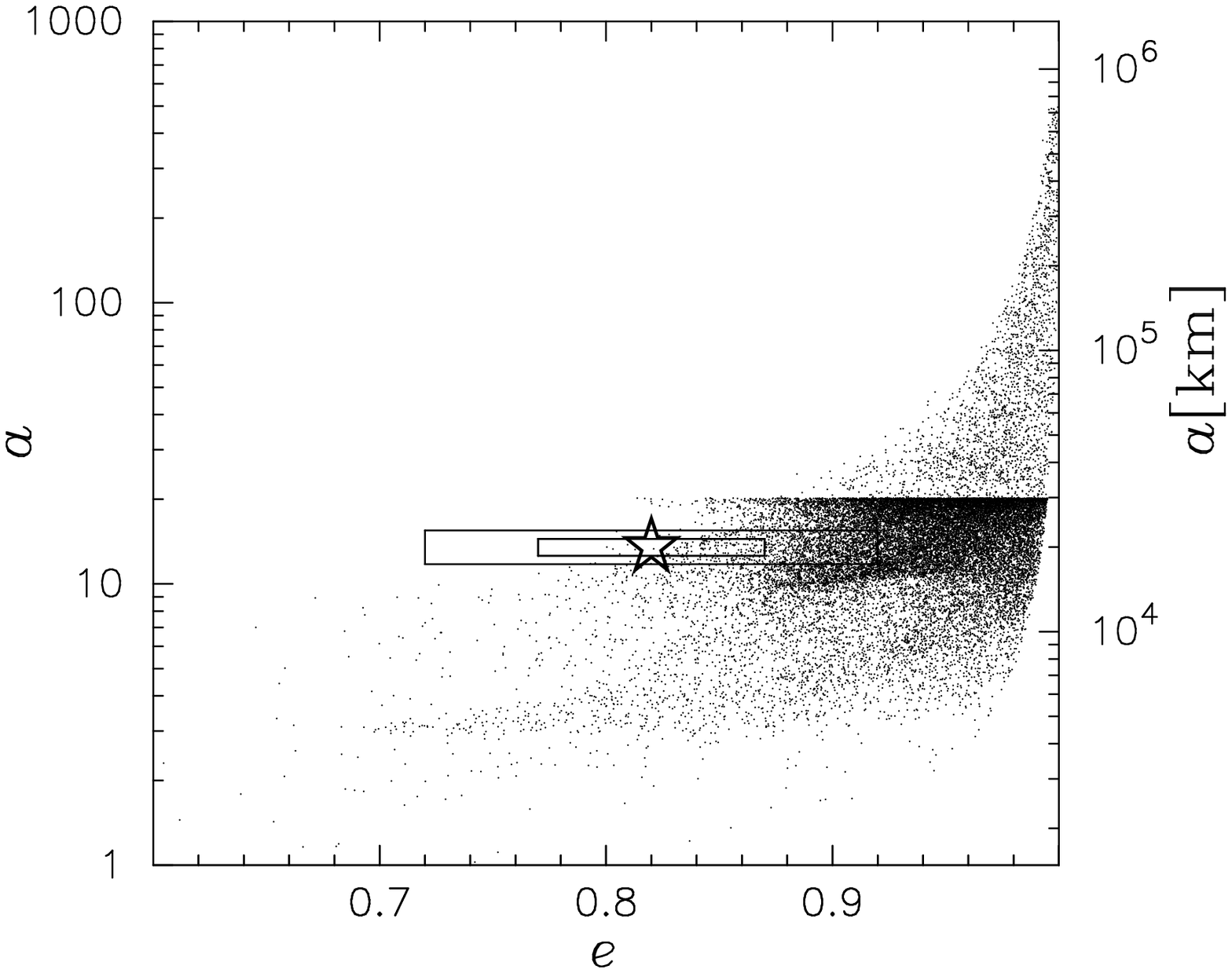}
\end{center}
\caption{Orbital properties of `massive-massive' binaries formed in
our scattering experiments: $a$ and $e$ have the same meaning and
units as in fig. 2.  Contributions from exchange reactions, channel (a)
in table 1, are limited by energy conservation to $a \simlt 20$,
and give rise to the horizontal rim in the middle of the figure.
Contributions involving mergers, channels (c) and (e) in table 1, can
lead to $a$ values all the way to the Hill radius $a \approx 10^3$, but
are limited by angular momentum conservation to increasingly high $e$
for increasing $a$.  The star symbol shows the observed orbit for
1998WW31.  Boxes around the star indicate the observational 1- and
2-$\sigma$ error bars.
}
\end{figure}

We are now in a position to confront our second task: to check whether
the exchange channel is efficient enough to produce the observed
binaries.  A straightforward approach would be to derive semi-analytic
estimates or to perform numerical simulations within the context of a
model for the protoplanetary disk.  Such an approach was taken by
Weidenschilling\cite{Weidenschilling2002} within the context of
simultaneous encounters of three massive bodies, and by Goldreich et
al.\cite{Goldreich2002} within the context of encounters of two
massive bodies while taking into account the dynamical friction from
the background sea of lighter bodies with or without the additional
passage of a third body.  In both cases, they had to assume local disk
densities high enough for their scenarios to work.

Interestingly, for our scenario we can simply bypass such an approach,
through the following observation.  Starting with TNOs of intermediate
mass, as predicted by Goldreich and Ward's theory for the formation of
planetesimals\cite{GoldreichWard1973}, the heaviest TNOs will accrete
mass primarily through collisions with other TNOs of comparable
mass\cite{KokuboIda1997, Makino1998}.
We also know that many of these collisions are of the `giant impact'
type that give rise to the formation of a tight circular strongly
unequal-mass binary as discussed above in channel 1.  We therefore
have to make no assumptions whatsoever about the density or other
physical parameters of the protoplanetary disk, since changing these
parameters will change the time scales for accretion and collisional
binary formation in exactly the same way.  All we have to do is
estimate what {\it fraction} of encounters between comparable-mass
TNOs will give rise to `giant impact' type binaries, and how long such
binaries survive on average before they are destroyed again.

For example, let us assume that one in three collisions between
comparable TNOs gives rise to a binary.  This is not unreasonable,
since in many collisions the encounter will be far from head-on, with
an angular momentum that would exceed that of the merger product even
if it would rotate at break-up speed, forcing the formation of an
accretion disk that is likely to form a small companion.  In two of
the three cases in which no binary is produced, we have to wait until
another major collision occurs.  Let us introduce this typical waiting
time as $T$.  What will happen, and how fast will it happen, in the
remaining one out of three cases in which we form a binary?

In the gravitational focusing regime, the cross section for
significant three body interactions to occur is
set by the size of the orbit $a_0$, while the cross section for direct
collisions is set by the size $r$ of the TNO primary.  Therefore, our
newly-formed binary is likely to undergo an exchange reaction on a
time scale $(r/a_0)T$, significantly smaller than $T$.  As a result, $a$
will increase significantly, as we have seen above.  Strong three-body
interactions will subsequently occur on an even much shorter time
scale $(r/a)T \ll T$.  After the first exchange reaction, a very wide
roughly equal mass binary exposed to encounters with a third body of
comparable mass will on average shrink the orbit and randomize the
eccentricity\cite{Heggie1975}.  As a result, the semimajor
axis will shrink systematically, while at any given time the
eccentricity will still be considerable, since the `thermal'
distribution $f(e) = 2e$ of eccentricities that follows phase space
volume is tilted towards high eccentricities.

When the orbit becomes small enough, subsequent three-body encounters
are more and more likely to lead to a physical collision between two
or three of the TNOs involved.  If all three collide, we are back
where we started, and the resulting system may be a single body (with
an assumed chance of 2/3) or a strongly unequal-mass binary (chance 1/3).
If two of the bodies collide, the third one may remain in orbit, with
a thermal ({\it e.g.} high) eccentricity expectation, or it may escape.
In the latter case, we again are back where we started.  In the former
case, we still have an equal-mass and likely highly eccentric binary.
Even its separation will still be high, for the following reason.
Long before the semimajor axis becomes comparable to the size of the
bodies, a typical three-body resonance reaction is likely to lead to a
collision between the three bodies, since each pass through the
resonant period gives a renewed chance for a collision.
The chance for collisions in resonant encounters becomes
significant\cite{HutInagaki1985}
when $r/a \sim 0.03$.  For simplicity, let us assume
that an exchange reaction turns a `giant impact' binary into a binary
with a semi-major axis of $a \sim 300r$.  Each subsequent strong
encounter will on average decrease $a$ by a factor\cite{HH2003} $\sim 1.2$.
We thus need to wait for a dozen such encounters to
occur before reaching $a \sim 30r$ and facing a significant chance for
a collision.  The time scale for each encounter to occur is $\sim (r/a)T$.
The waiting time for the last encounter in this series to occur, under
these simplifying assumptions, is $(1/30)T$, while each previous
waiting time was less by a factor 1.2.  Summing this series, we get a
total waiting time of $(T/30)(1 + (1/1.2) + (1/1.2)^2 + \dots)\approx
(T/30)/(1 - (1/1.2)) = 0.2T$, as an estimate for the duration for
those dozen encounters to happen.

Under these assumptions, in 1/3 of the cases, we wind up with an
equal-mass TNO binary with the observed properties for a period $\sim 0.2T$,
compared to a 2/3 chance to wind up with a single TNO for a period
$\sim T$.
This allows us to derive the rate equation for the formation and
destruction of the binaries. 
If we denote by $N_S$ and $N_B$ the number of single bodies and the
number of binaries, respectively, we have
\begin{eqnarray}
\frac{dN_B}{dt} &=& \frac{1}{3}N_S \, -
                    \, \frac{1}{0.2} \,\frac{2}{3}N_B\nonumber\\
\frac{dN_S}{dt} &=& -\,\frac{dN_B}{dt}\nonumber
\end{eqnarray}
If we measure time in unit of $T$. So for the stationary state we have
$dN_B/dt = dN_S/dt=0$, and $N_B=0.2N_S/2=0.1N_S$. Therefore, the
binary fraction is $\sim 10$\%.  When accretion in the Kuiper belt
region diminished, the number of single and binary objects was frozen,
with a ratio similar to this steady-state value.

Clearly, this predicted binary fraction of 10\% is dependent on the
assumption we made for the size of a newly formed equal-mass binary,
and our other estimates have also been rather crude.  It is clear,
however, that we expect to find {\it any} TNO as part of a very
wide eccentric equal-mass binary during at least a few percent of its
history.  This implies that among all TNOs, after cessation of the
accretion stage several percent or more were accidentally left in such a
binary phase.  The fact that more than 1\% of the known TNOs are found
to be in wide roughly equal-mass binaries is thus a natural
consequence of {\it any} accretion model {\it independent of the
assumed parameters} for the density and velocity dispersion of the
protoplanetary disk or the duration of the accretion phase.

We conclude that we have found a robust and in fact unavoidable way to
produce the type of TNO binaries that have been found.  As a
corollary, we predict that future discoveries of TNO binaries will
similarly show roughly equal masses, large separations, and high
eccentricities.

{\bf Acknowledgements}
We acknowledge helpful comments on our manuscript by Peter Goldreich
and Roman Rafikov.

\newpage

\end{document}